# A Novel ML-Fuzzy Control System for Optimizing PHEV Fuel Efficiency and Extending Electric Range under Diverse Driving Conditions

Mehrdad Raeesi[1*], Saba Mansour[2], Sina Changizian[3]

[1] School of Automotive Engineering, Iran University of Science and Technology, P. O. Box 1681666110, Tehran, Iran
[2] Sibley School of Mechanical and Aerospace Engineering, Cornell University, Ithaca, NY 14850, United States
[3] College of Engineering, Mechanical and Material Department, University of Nebraska-Lincoln, Lincoln, United States
[*] Corresponding author: Mehrdad Raeesi, mehrdadraeesi@gmail.com

## Abstract

Aiming for a greener transportation future, this study introduces an innovative control system for plug-in hybrid electric vehicles (PHEVs) that utilizes machine learning (ML) techniques to forecast energy usage in the pure electric mode of the vehicle and optimize power allocation across different operational modes, including pure electric, series hybrid, parallel hybrid, and internal combustion operation. The fuzzy logic decision-making process governs the vehicle control system. The performance was assessed under various driving conditions. Key findings include a significant enhancement in pure electric mode efficiency, achieving an extended full-electric range of approximately 84 kilometers on an 80% utilization of a 20-kWh battery pack. During the WLTC driving cycle, the control system reduced fuel consumption to 2.86 L/100km, representing a 20% reduction in gasoline-equivalent fuel consumption. Evaluations of vehicle performance at discrete driving speeds, highlighted effective energy management, with the vehicle battery charging at lower speeds and discharging at higher speeds, showing optimized energy recovery and consumption strategies. Initial battery charge levels notably influenced vehicle performance. A 90% initial charge enabled prolonged all-electric operation, minimizing fuel consumption to 2 L/100km less than that of the base control system. Real-world driving pattern analysis revealed significant variations, with shorter, slower cycles requiring lower fuel consumption due to prioritized electric propulsion, while longer, faster cycles increased internal combustion engine usage. The control system also adapted to different battery state of health (SOH) conditions, with higher SOH facilitating extended electric mode usage, reducing total fuel consumption by up to 2.87 L/100km.

## 1. Introduction

The transportation sector is undergoing a revolution with the rise of electric vehicles (EVs). These vehicles promise a cleaner and more sustainable future by addressing energy security concerns and reducing greenhouse gas emissions. A key solution lies in increasing energy conversion efficiency and reducing emissions, and plug-in hybrid electric vehicles (PHEVs) are particularly attractive. PHEVs address a major barrier for pure electric vehicles, i.e., range anxiety. They offer a compromise by combining an internal combustion engine (ICE) with an electric motor powered by a rechargeable battery. Drivers can utilize either gasoline or renewable fuels for the ICE, while the electric motor provides a zero-emission option.

This hybrid approach promotes energy diversification and efficiency by offering drivers versatility based on their needs.

However, optimizing power distribution and management in PHEVs remains a challenge [1]. The complex interplay between the ICE, electric motor, battery, and real-time driving conditions necessitates sophisticated energy management strategies (EMSs) [2]. These strategies aim to find the optimal balance between power sources, maximizing fuel economy and minimizing emissions while preserving battery health through maintaining the desired state of charge (SOC). Optimizing drivetrain components poses a significant challenge in plug-in hybrid electric vehicle technology [3–5]. For instance, Song et al. [6] developed a simulation platform tailored for optimizing PHEVs in bus applications. Their multi-objective approach yielded promising results, notably reducing fuel consumption and enhancing performance. Similarly, Ribau et al. [7] utilized optimization techniques to enhance the performance and cost-effectiveness of both hybrid electric vehicles (HEVs) and PHEVs. Qi et al. [8] introduced an online energy management system for PHEVs, leveraging evolutionary algorithms to minimize powertrain costs while meeting driver requirements, demonstrating robustness in real-world traffic conditions.

Regarding component sizing, Xu et al.[4] proposed a model for optimizing fuel cell and battery sizes in PHEV buses, revealing the impact of factors like fuel cell efficiency and battery capacity on vehicle range and performance. Redelbach et al. [3] underscored the significance of battery size for PHEV ownership costs, suggesting modular designs to cater to individual needs. Hu et al. [7] investigated the effects of smaller batteries on PHEV buses, in another study, Raeesi et al. [9] evaluated the impact of fuel cell degradation on the performance and power distribution of fuel cell vehicles (FCVs) across varied driving conditions, emphasizing the need to assess the effects of increased fuel cell consumption on vehicle performance and battery efficiency to enhance fuel consumption. These endeavors highlight the intricate challenges of optimizing PHEVs for real-world driving scenarios. While notable progress has been achieved in component sizing and control strategies, further research is required to address the complexities of balancing efficiency, emissions, and cost-effectiveness across diverse driving contexts [5].

Power management strategies for PHEVs have largely evolved from those utilized for conventional hybrids, falling into two main categories: rule-based and optimization-based approaches [10–13]. Rule-based strategies distribute power within the vehicle solely based on its current state (such as vehicle/engine velocity and SOC) and input variables (power demand), employing predefined rule maps like deterministic or fuzzy controllers [14–16]. While being relatively straightforward to implement and ensuring close-to-optimal operation of power suppliers, rule-based strategies lack global optimality due to variations in driving behavior that deviate from predefined cycles or maps [17].

On the other hand, optimization-based strategies seek to achieve global optimization by determining optimal control actions without necessarily requiring knowledge of future power demand. Methods such as deterministic dynamic programming, quadratic programming, and neural networks are employed when the driving cycle is known beforehand [18–20]. Various optimization techniques, including sliding mode control and particle swarm optimization, are utilized to optimize powertrain parameters and control strategies [21,22]. Additionally, adaptive equivalent consumption minimization strategies and predictive techniques utilizing GPS (global positioning systems), GIS (geographic information systems), and traffic flow modeling aim to optimize real-time energy management [23–25]. However, these approaches face challenges when driving cycles become dynamic and uncertain due to factors like varying driving speeds and commuting times influenced by individual habits and real-time traffic conditions. Such uncertainties pose significant challenges in optimizing fuel consumption and emissions in PHEVs, emphasizing the need for robust and adaptive control strategies.

Stochastic models have emerged to address uncertainties in driving cycles. These models optimize power management based on probabilities rather than single, predefined cycles. However, existing approaches often rely on the assumption of infinite commuting time, requiring a discount factor in optimization algorithms [26,27]. The selection criteria used for this factor in the literature are unclear, potentially impacting results and convergence rates. A prevalent approach in HEV energy management is the equivalent consumption minimization strategy (ECMS) [28]. ECMS formulates the global objective of minimizing fuel consumption as a local optimization problem. A critical parameter in ECMS is the equivalence factor (EF), which determines the power distribution between the engine and battery. However, existing methods for determining the optimal EF often rely on offline calculations and struggle to adapt to real-world driving variability [28–30].

This limited adaptability to real-time conditions necessitates online EF adjustment strategies that consider factors like driving patterns and battery state of charge [31]. Research suggests promising approaches using velocity prediction, driving pattern recognition, and battery state-of-charge feedback [31]. Recent research has explored online EF adaptation strategies. Han et al. [18] proposed an EF adaptation rule based on velocity prediction using a neural network, achieving fuel economy improvements of 2.7% to 7% in simulations. Other researchers have focused on feedback-based adaptation without future prediction, leveraging real-time battery SOC for improved optimization [19, 20]. Yang et al. [19] developed a SOC-dependent EF map and online PI (proportional integral) control for PHEV city buses, demonstrating a 15.93% fuel economy improvement.

Deep learning (DL) neural networks have emerged as powerful tools for various engineering problems, including energy systems [32]. This has led to the development of data-driven energy management systems for PHEVs, such as the artificial neural network ECMS (ANN-ECMSs), that utilize real-world driving data to optimize performance [33]. However, such approaches often rely on simplified driving cycles, limiting their effectiveness in real-world scenarios with unpredictable driving patterns [34]. A well-designed EMS offers benefits beyond just fuel economy. It can also extend battery life and capacity by optimizing charging and discharging cycles [35]. Research has shown that the state of charge is a key factor in battery degradation, with temperature playing a secondary but significant role [36]. Optimizing such factors through EMS can significantly improve battery health, which in turn has a great impact on the driving range [37].

Life-cycle assessments (LCAs) are crucial for evaluating the environmental and economic benefits of PHEVs [38]. These studies consider not only the production phase of lithium-ion batteries but also their usage and potential for recycling. Additionally, LCA can assess the life cycle cost and greenhouse gas emissions of PHEVs compared to traditional vehicles under various driving scenarios [39].

Following a detailed examination of the literature, it becomes evident that energy management and enhancing the efficiency of PHEVs are key challenges. Effective energy management directly impacts pollutant emissions. In light of this, this study aims to address these challenges by employing an intelligent approach to model the longitudinal dynamic performance of PHEVs using fuzzy control. The methodology involves several steps:

- Collection and analysis of real-world driving cycles in Tehran.

- Detailed simulation of vehicle components, focusing on the battery and engine models, using Simcenter AMESim software.
- Development of an intelligent model to predict the future charging level of the vehicle based on driving parameters.
- Integrating the battery consumption prediction model with fuzzy logic to devise a control system for power allocation and vehicle performance across various modes: full-electric, series, parallel, or full ICE.

Through these steps, we aim to evaluate the performance of electric vehicles under different driving conditions while considering the battery state of health. This comprehensive approach seeks to enhance the PHEVs' efficiency and environmental impact in real-world scenarios.

## 2. Modeling, topology, and governing equations

This subsection details the vehicle powertrain components and modeling methodology within the Simcenter Amesim software. The primary focus is on a plug-in hybrid electric vehicle with a series-parallel configuration, capable of operating in series, parallel, all-electric, and internal combustion engine modes. The modeled PHEV prioritizes all-electric operation, resulting in a larger battery pack compared to conventional hybrids in the same class. Consequently, the vehicle weight is higher than similar ICE models. Even so, as shown in Table 1, the PHEV achieves superior fuel efficiency across all modes compared to both series and parallel hybrids, as well as conventional ICE vehicles.

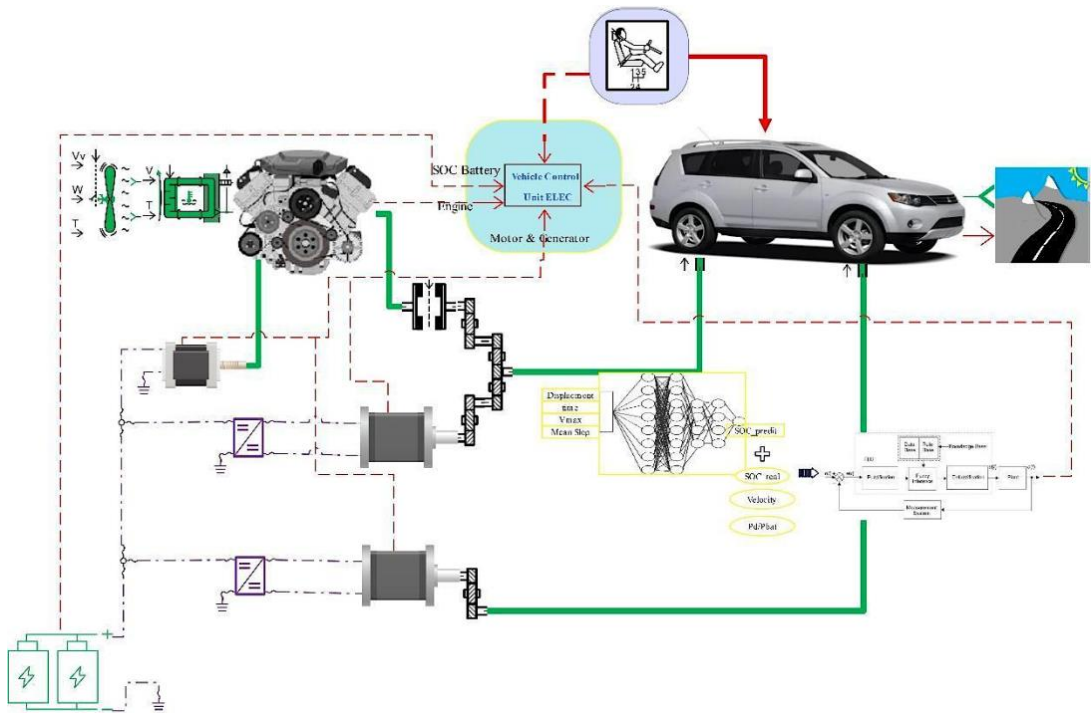

Figure 1: Schematic of the developed model in Simcenter Amesim

The schematic representation of the model developed in Simcenter Amesim is shown in Figure 1. The interactions between the key components, including two electric motors, a generator, and a 2.4-liter ICE engine with a maximum power output of 99 kW are illustrated in this figure. Further details of the performance specifications of these components are summarized in Table 1.

Table 1- Mitsubishi Outlander PHEV parameters [40], [41]

| Component | Specification | Value/Type | Unit |
| --- | --- | --- | --- |
| | Total displacement | 2360 | cc |
| Engine | Maximum output | 99 / 4500 | kW/rpm |
| | Maximum torque | 211 / 4500 | N.m/rpm |
| Fuel system | Fuel tank capacity | 43 | L |
| | Rated output | 25 | kW |
| Front electric motor | Maximum output | 60 | kW |
| | Maximum torque | 137 | N.m |
| | Rated output | 30 | kW |
| Rear electric motor | Maximum output | 70 | kW |
| | Maximum torque | 195 | N.m |
| | Cell type | Lithium-ion | - |
| Battery | Maximum operating voltage | 300 | V |
| | Capacity | 13.8 | kWh |
| | 100% with 13 A charger | 300 | minutes |
| Charging times | 100% with 16 A charger | 210 | minutes |
| | Rapid charge to 80% | 25 | minutes |

### Longitudinal driver

To understand and evaluate the vehicle acceleration and deceleration capabilities, a longitudinal dynamic model is developed. This mathematical model represents the vehicle's longitudinal motion along its forward direction. Vehicle speed alterations are considered by calculating the forces and moments acting on the vehicle during driving cycles. The engine torque output, transmission gear ratio, tire characteristics, and aerodynamic drag are all factors that should be considered [46]. The forces that impact the vehicle's forward movement play a central role in the longitudinal dynamics model. Other important factors are:

- Traction force: Generated by the engine and transmitted through the driveshaft to the wheels, propelling the vehicle forward.
- Tire interaction: Representing the complex forces between the tires and the road surface, affecting factors like grip and rolling resistance.
- Aerodynamic drag force ($F_{aero}$): Air resistance acting against the forward movement of the vehicle, increasing with speed.
- Rolling resistance ($F_{roll}$): Resulting in energy loss caused by the deformation of the tires as they roll on the road.

The model assumes a state of force equilibrium. This means that the sum of all longitudinal forces acting on the vehicle must equal its mass multiplied by its acceleration. This principle is expressed as:

$$F_{tot} = m \times a, \tag{1}$$

where $F_{tot}$ is the sum of the forces, $m$ is the total vehicle mass, and $a$ is the acceleration.

The position of the pedal determines the acceleration or deceleration modes. The velocity of the present moment is analyzed and compared to the velocity determined by the input data at each time step. The intended mode is to stop if the vehicle speed at the moment ahead is higher than the set speed (with a slight delay). Otherwise, it is the acceleration mode, i.e., applying force to the gas pedal. Consequently, the force is calculated using [3].

$$F_{cl} = (m + load) \times g \times sin\left(arctan\frac{\alpha}{100}\right), \tag{2}$$

where $F_{cl}$ is the traction force, measured in Newtons (N), $m$ is the mass of the vehicle in kg, $load$ is characteristic of the additional load of the car in kg, $g$ is the acceleration of gravity in $m/s^2$, and the slope of the road, which is entered as a percentage, is indicated by the alpha attribute. Force per constant speed of the car, in Newtons (N), can be calculated as:

$$F_{stab} = A.(V_c > 0) + B.V_c + C.V_c^2 + F_{cl}, \tag{3}$$

in which, the vehicle speed is represented by $V_c$ in $m/s$, and the rest of the parameters are constant coefficients. When the car is traveling on the road, the incoming forces can be calculated as [42]:

$$F_{aero} = \frac{1}{2}.\rho_{air}.S.C_x.(V_c + V_w)^2, \tag{4}$$

$$F_{roll} = (m + load).g.(f + k.(V_c + V_w) + w.(V_c + V_w)^2), \tag{5}$$

$$F_{stab} = F_{aero} + F_{roll} + F_{cl}. \tag{6}$$

Wherein the aerodynamic and rolling friction forces, $F_{aero}$ and $F_{roll}$, have units of N, $\rho_{air}$ represents the air density parameter in $kg/m^3$, $S$ represents the area in front of the car in $m^2$, $C_x$ is the drag coefficient constant, and $V_w$ is the wind speed in $m/s$. The total traction force is:

$$F_{tot} = a \times m_{eq} + F_{stab}, \tag{7}$$

which results in a total power of [43]:

$$Power = F_{tot} \times V_c. \tag{8}$$

### Internal combustion engine

Energy losses in an internal combustion engine encompass several processes, including combustion, heat dissipation, pumping losses, and power required to operate auxiliary equipment such as the alternator, air conditioning, and power steering systems. Modeling these processes with physical equations under driving conditions demands significant computational resources. Hence, a general summary of the engine properties is necessary. This information can be presented in various ways to quantify fuel consumption under different conditions, such as varying engine speeds, torques, and throttle positions. The most common technique to summarize overall engine fuel efficiency is the brake specific fuel consumption (BSFC) map.

To determine the BSFC of an engine under various operational conditions, empirical measurements or combustion simulations are used. The BSFC is calculated as:

$$bsfc = \frac{3600\ f}{P_b} \tag{9}$$

where $f$ is the fuel consumption rate in grams per second (g/s) and $P_b$ is the power output in kilowatts (kW). Additionally, the engine efficiency can be expressed in terms of BSFC as:

$$\eta_f = \frac{3.6 \times 10^6}{bsfc \times h_u} \tag{10}$$

where $h_u$ is the fuel's energy content in joules per gram (J/g). In PHEVs, part of the required power is supplied by the engine. Denoting this power as $P_e$, the total fuel consumption $FC_t$ over time can be calculated as follows:

$$FC_t = \sum_{i}^{N} P_{e,i} \times \frac{bsfc_i}{1000\gamma_f} \Delta t_i \tag{11}$$

where $\gamma_f$ is the fuel density in kilograms per liter ($kg/L$).

### Emissions

The emission levels of nitrogen oxides (NOx), carbon monoxide (CO), unburned hydrocarbons (HC), and particulate matter are critical performance characteristics of an engine. The concentrations of these gaseous pollutants in the engine exhaust are usually measured in parts per million (ppm) or percent by volume (mole fraction). Similar to BSFC, these emissions are often evaluated based on the engine power output. Thus, to define the specific emission parameters, we can have:

$$SE_x = \frac{3600 \times \dot{m}_x}{P_e} \tag{12}$$

where $x$ represents the specific emission mentioned. The total specific emissions $SE_x$ can be determined over time by summing the values in all time steps. This approach provides a comprehensive understanding of fuel consumption and emissions, essential for optimizing engine performance and meeting regulatory standards.

### Battery modeling

To simulate the operational characteristics of the battery pack, the equivalent circuit model (ECM) can be employed as a mathematical framework [5]. This model comprises voltage sources, resistors, and capacitors, which are accurately interconnected within the electronic control module. Empirical data is utilized to calibrate the ECM, determining the values of its components [6]. The ECM can predict battery pack performance under varying discharge rates, temperatures, and states of charge (SOC). This predictive capability is valuable for designing battery systems tailored to specific applications, such as electric transportation or industrial equipment. Various electronic control mechanisms can simulate battery packs, with the assumption that all cells are identical. The equivalent circuit model, illustrated in Figure 2, represents all the cells, the additional resistance from connections, and other internal components within the battery pack.

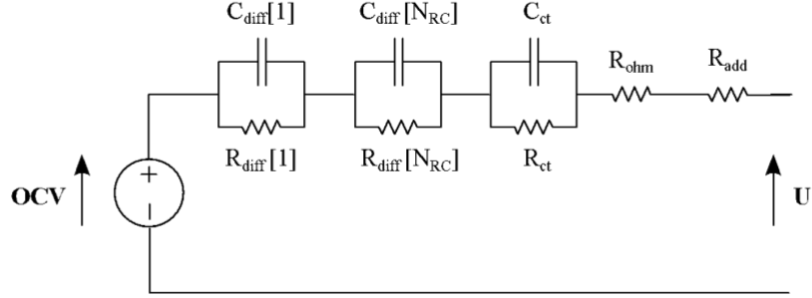

Figure 2: Schematic of the equivalent circuit model of the battery assembly

The additional resistance ($R_{add}$) depends on the number of cells arranged in parallel and series, as well as other parameters derived from the cell characteristics. The state of charge (SOC) is calculated using the following equation [44]:

$$d(SOC)/dt = 100 \times I/Q' \times \eta_{farad}, \tag{13}$$

where $I$ represents the battery current in amperes (A), $Q'$ is the available battery capacity in ampere-seconds (A.s), and $\eta_{farad}$ is Faraday efficiency. The voltage drops resulting from additional resistance, ohmic resistance, and load transfer are respectively given by:

$$\Delta U_{add} = -I \times R_{add}, \tag{14}$$

$$\Delta U_{ohm} = -I \times R_{ohm}, \tag{15}$$

$$\Delta U_{ct} = -I \times R_{ct}. \tag{16}$$

The amount of residual voltage drop is equal to

$$\Delta U_{hyst} = OCV_{eq} - OCV, \tag{17}$$

and the amount of voltage drop caused by diffusion is

$$\Delta U_{diff}^{total} = \sum_{i=1}^{i=N_{RC}} \Delta U_{diff}[i], \tag{18}$$

which yields the amount of total voltage drop

$$\Delta U_{total} = \Delta U_{add} + \Delta U_{ohm} + \Delta U_{ct} + \Delta U_{hyst} + \Delta U_{diff}^{total}. \tag{19}$$

## Equations of the battery thermal model

Dissipative heat in the battery pack includes additional resistance loss ($Q_{add}$), Faraday loss ($Q_{\eta_{farad}}$), entropic heat flow ($Q_{\frac{dU}{dT}}$), hysteresis loss ($Q_{hyst}$), ohmic resistance loss ($Q_{ohm}$), charge transfer loss ($Q_{ct}$), and diffusion loss ($Q_{diff}$). These losses are calculated through [45]

$$Q_{add} = -I \times \Delta U_{add}, \tag{20}$$

$$Q_{\eta_{farad}} = I \times OCV_{eq} \times (1 - \eta_{farad}), \tag{21}$$

$$Q_{\frac{dU}{dT}} = -I \times \frac{dU}{dT} \times (T + 273.15) \times \eta_{farad}, \tag{22}$$

$$Q_{hyst} = -I \times \Delta U_{hyst}, \tag{23}$$

$$Q_{ohm} = -I \times \Delta U_{ohm}, \tag{24}$$

$$Q_{ct} = -I \times \Delta U_{ct}, \tag{25}$$

$$Q_{diff} = -I \times \Delta U_{diff}^{total}. \tag{26}$$

Therefore, the total heat loss can be written as

$$Q = Q_{add} + Q_{\eta_{farad}} + Q_{\frac{dU}{dT}} + Q_{hyst} + Q_{ohm} + Q_{ct} + Q_{diff}. \tag{27}$$

More complex ECMs, such as two-node and three-node models, can be used to get more accurate predictions of the performance of a battery pack. These models include additional components, such as capacitors, that can represent the effects of battery aging and other factors. The accuracy of an ECM depends on several factors, including the type of model, the quality of the experimental data used to fit the model, and the specific application for which the model is being used. More complex models are generally more accurate, nonetheless, this extra accuracy comes at the cost of the model being more difficult to use.

## Lithium-ion battery degradation

Lithium-ion batteries experience a gradual decline in state of health (SOH) as they undergo repeated charging and discharging cycles [46]. This degradation is a combination of mechanical and chemical processes. As for mechanical degradation, the volume changes during charging and discharging cycles can cause physical stress on the battery, leading to wear and tear [47]. On the other hand, chemical degradation happens because chemical reactions within the battery, like electrolyte breakdown and the formation of the solid-electrolyte interface (SEI), contribute to lithium-ion loss and increased internal resistance [48]. These combined effects can manifest as reduced battery capacity and higher internal resistance. The battery typically reaches its end-of-life (EOL) when one of two following criteria is met; Either internal resistance increases by 100% or capacity decreases by 20% [40]. Since reaching EOL signifies the need for battery replacement, determining lithium-ion battery health is crucial. Two common methods of assessing SOH rely on comparing current performance with the initial properties of the battery [49].

The first method is the capacity-based determination of SOH. This method compares the battery's current capacity ($C_a$) to its initial rated capacity ($C_{rated}$). The SOH, in percentage, is thus expressed as:

$$SOH = \frac{C_a}{C_{rated}} \times 100. \tag{28}$$

The second method is the resistance-based evaluation of SOH. This method compares the battery's current internal resistance ($R_{cur}$) to its resistance at end-of-life ($R_{EOL}$) and new state ($R_{new}$). Like in the previous method, the SOH is expressed as a percentage

$$SOH = \frac{R_{EOL} - R_{cur}}{R_{EOL} - R_{new}} \times 100. \tag{29}$$

Although direct measurement of capacity and internal resistance with current commercial sensors is not feasible, it is imperative to estimate the SOH. Hence, experts approximate these values by considering quantifiable parameters like voltage, current, and temperature. The aim of this field is to actively seek out reliable and efficient methods for assessing the safety of various battery designs and diagnosing faults.

## 3. Driving cycle

After examining performance metrics and governing equations, the crucial concept of driving cycles should be examined. These cycles serve as virtual roadmaps, capturing vehicle behavior and driving conditions over a specific distance. Driving cycles are typically visualized as a time-speed diagram to mimic real-world driving patterns. Simulations of real-world driving cycles, such as those performed in this study, encompass various vehicle states, including acceleration, deceleration, constant speed, and idling. Different driving cycles are considered in this study, including the worldwide coordinated light vehicle test cycle (WLTC), and four real-world driving profiles from the city of Tehran, each representing different driving behaviors. These real-world profiles are compared to the WLTC cycle in Table 2.

Table 2: Statistical characteristics related to different studied driving cycles, i.e., routes

| Parameter | WLTC | Tehran 1 | Tehran 2 | Tehran 3 | Tehran 4 | Tehran 5 | Tehran 6 | Unit |
|---|---|---|---|---|---|---|---|---|
| Total distance | 23450 | 66714.34 | 17895.76 | 34096.90 | 128731.56 | 40936.84 | 46571.59 | m |
| Total time | 1800 | 1822.00 | 2588.00 | 1975.00 | 4485.00 | 969.00 | 1468.00 | s |
| Driving time | 1573.4 | 1735.00 | 2270.00 | 1834.00 | 2887.00 | 931.00 | 1368.00 | s |
| Stop time | 250 | 87.00 | 318.00 | 141.00 | 1598.00 | 38.00 | 100.00 | s |
| Average speed | 46 | 38.45 | 7.88 | 18.59 | 44.59 | 43.97 | 34.04 | km/h |
| Max speed | 131.2 | 86.39 | 72.67 | 53.23 | 94.15 | 98.30 | 79.81 | km/h |
| Number of stops | 8 | 2 | 25 | 4 | 3 | 1 | 3 | - |

A key benefit of real-world driving cycles is incorporating actual road conditions, including altitude that allows us to estimate road slope, providing a more realistic vehicle performance compared to standardized cycles. Since driving cycles play a crucial role in assessing vehicle performance and emissions, real driving cycles can be beneficial.

## 4. Control strategy

The increasing demand for sustainable transportation has motivated the development of PHEVs, integrating electric and gasoline engines to enhance fuel efficiency and range. Nonetheless, optimizing the power management of PHEV powertrains to minimize fuel consumption and emissions poses a multifaceted challenge. Conventional control methods often rely on predefined rules and thresholds, which may not effectively adapt to real-time driving conditions. In this study, we propose a novel data-driven approach coupled with fuzzy logic control systems for PHEVs, addressing the shortcomings of traditional methodologies. Figure 3 exhibits the proposed control model development flowchart, where DOE stands for design of experiment. This innovative system comprises three core components:
1. Battery state of charge estimation: Implementing a deep learning model trained based on extracted data from standard real-world driving cycles, resulting in accurate estimation of PHEV battery SOC for the all-electric operating mode. This estimation is pivotal for determining vehicle performance mode and optimizing engine management.
2. Operating mode detection: Utilizing SOC estimation, real-time sensor data, and a data-driven methodology, the system identifies speed levels, current charge levels, and requested power. These parameters serve as inputs to the fuzzy control system for estimating functional modes.
3. Fuzzy logic-based propulsion management: Employing a fuzzy logic system that emulates human reasoning, the estimated SOC, operating modes, and other relevant parameters are utilized to ascertain the optimal power distribution between the electric motor and the internal combustion engine. This intelligent control strategy ensures efficient PHEV operation, minimizing fuel consumption and emissions.

By integrating these components, our proposed controller strategy offers a robust and adaptive solution for managing PHEV power flow, enabling enhanced performance under varying driving conditions while promoting sustainability in transportation.

The control strategy aims to achieve three main objectives:
1. Maximizing the electric driving range: The system prioritizes electric driving whenever possible to maximize the PHEV's electric range and minimize fuel consumption.
2. Minimizing emissions: The system minimizes emissions by reducing the use of the internal combustion engine when possible.
3. Satisfying power demand: The system ensures that the PHEV meets the driver's power demand by effectively managing the power split between the electric motor and the internal combustion engine.

To achieve these objectives, the fuzzy logic system considers various factors, including:
- Estimated SOC: The SOC indicates the remaining battery capacity, influencing the decision to switch to electric or hybrid modes.

- Operating mode: The current operating mode determines the allowable power range for the electric motor and the internal combustion engine.
- Vehicle speed and acceleration: These parameters indicate the power demand, influencing the power split between the electric motor and the internal combustion engine.
- Driver power request: The driver's power request is considered to ensure that the PHEV responds to the driver's input.

By dynamically adjusting the power split based on these factors, the fuzzy logic system optimizes PHEV performance, achieving the desired balance between fuel efficiency, emissions, and power delivery.

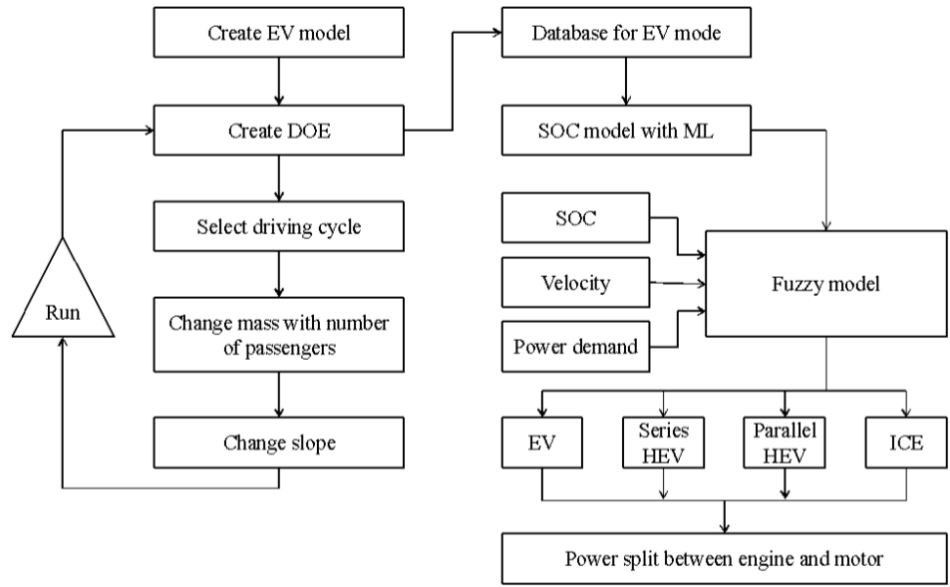

Figure 3: Control model development diagram: Integrating machine learning and fuzzy logic

### Predicted SOC for EV mode

As illustrated in Figure 3, for the fuzzy control system to make more accurate decisions, it is essential to input the predictive parameter of the battery charge level for the all-electric operation mode into the system. This section details the method used to extract the relevant database, as shown in Figure 3. A machine learning model, developed using the TensorFlow framework, is incorporated to enhance the decision-making process. The deviations of the predicted data for both the training and test datasets are shown in Figure 4. As shown in Figure 4, the closer the fidelity value ($R^2$) to one, the more accuracy in the predictive model.

### Fuzzy logic control system for PHEV operating modes

This paper introduces a fuzzy logic control system designed to identify the optimal operating mode for a PHEV. Unlike traditional control strategies that depend on fixed rules and thresholds, which may not adequately respond to real-time driving conditions, fuzzy logic offers a more adaptable approach. By incorporating partial membership within categories, fuzzy logic allows for nuanced decision-making, thereby enabling the system to consider a broader variation of operating conditions. Where partial membership stands for a degree of membership ranging from 0 (no membership) to 1 (full membership), rather than having a binary membership status.

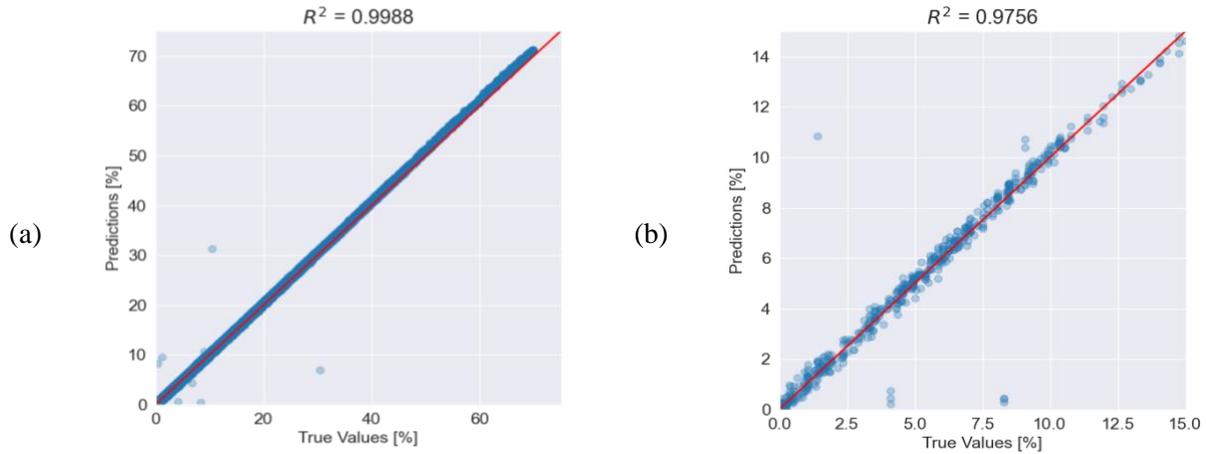

Figure 4: Deviation of predicted data for a) training and b) test datasets

### System inputs and outputs

The system utilizes several key inputs to determine the optimal operating mode:

- Vehicle speed (km/h): Represents the current speed of the vehicle.
- State of charge (SOC) (%): Indicates the remaining battery capacity.
- Predicted state of charge ($SOC_{pred}$) (%): Represents the estimated future battery level based on a deep learning model.
- Required power (kW): Represents the power demand from the driver.

The output of the system is the operating mode, which determines the most suitable powertrain configuration for the current driving scenario. The operating modes considered are:

- Electric vehicle: The vehicle operates solely on the electric motor, maximizing fuel efficiency and minimizing emissions.
- Parallel hybrid: Both the electric motor and internal combustion engine (ICE) contribute to propulsion, offering a balanced situation between fuel efficiency and power delivery.
- Series hybrid: The ICE acts as a generator to power the electric motor, which drives the wheels. This mode is often utilized for extended electric range when SOC is low.
- Combustion engine: The vehicle operates solely on the ICE, typically used when battery capacity is depleted or high-power demands are required.

### Fuzzy membership functions and rules

The system defines fuzzy membership functions for each input and output variable. These functions represent the degree of membership of a specific value within a category (e.g., low, medium, or high). Triangular membership functions are commonly used to represent the gradual transition between categories.

# Rule development and control objectives

The fuzzy logic control system utilizes a set of 30 rules to determine the optimal operating mode for the PHEV. These rules are designed to achieve several key objectives, which can be broadly categorized into the following areas:

**1.      Prioritizing electric driving**

A primary goal of the control system is to maximize the use of electric propulsion, which contributes to fuel efficiency and emission reduction. This is achieved by favoring EV mode under conditions where electric operation is feasible and efficient.

- **High SOC and low power demand:** When the battery has a sufficient charge (high SOC) and the required power is low, the system prioritizes EV mode to utilize the electric motor directly. This is particularly beneficial in low-speed scenarios (urban conditions) where electric propulsion is most efficient.

**2.      Managing battery depletion**

The control system also considers the predicted SOC alongside the current SOC to proactively manage battery depletion. When both SOC and predicted SOC are low, the system prioritizes modes that conserve battery life and prevent premature depletion.

- **Low SOC and high predicted depletion**: If the current SOC is low ($SOC < 30$) and the predicted SOC indicates further depletion, the system may switch to combustion engine mode or series hybrid mode. This mode prevents the battery from reaching critically low levels and ensures sufficient power availability.
- **Low SOC at highway speeds**: At highway speeds, where the output of the electric motor may not be efficient, the system may favor series hybrid mode over EV mode even with a low SOC. This mode allows the combustion engine to propel the vehicle completely.

**3.      Optimizing powertrain efficiency**

The control system balances battery utilization with powertrain efficiency by considering both SOC and required power. When SOC is high ($SOC > 80$) or medium ($30 < SOC < 80$) and power demand varies, the system may favor parallel hybrid mode.

- **High/medium SOC and varying power demand:** In situations with high or medium SOC and varying power demands (low to high), the parallel hybrid mode allows the system to combine electric motor and combustion engine operation for optimal efficiency. This leverages the strengths of both engines across a range of power requirements.
- **Low SOC and high power demand:** Even with low SOC, the system may favor parallel hybrid mode for high power demands. This ensures sufficient power delivery by utilizing both the electric motor and combustion engine, even when the battery charge is not ideal.

4. **Maintaining power delivery**

The control system prioritizes power delivery by ensuring sufficient torque and power are available to the wheels under various driving conditions.

- **High Power Demand:** When the required power exceeds the maximum battery power, the system prioritizes parallel hybrid mode, leveraging both the electric motor and the combustion engine. This approach ensures that the PHEV can meet the driver's power demands, even when the battery charge is suboptimal.

5. **Regenerative braking**

The control system utilizes a regenerative braking system to capture energy during braking and recharge the battery.

- **Braking:** During braking, when the required power is negative, the system prioritizes EV mode. This allows the electric motor to act as a generator, capturing energy from the wheels and recharging the battery.

By incorporating these objectives into the fuzzy rules, the control system strives to achieve a balance between fuel efficiency, emission reduction, power delivery, and battery management for optimal PHEV operation under various driving conditions. The system adapts its decision-making mode based on real-time inputs, ensuring a dynamic and responsive approach to PHEV control.

# 5. Results and discussions

In this section, the performance of the proposed control system for an electric vehicle is evaluated under various driving conditions. The assessment includes three stages:

- Evaluation in All-Electric Mode: The vehicle performance in all-electric mode is evaluated, examining its energy consumption, driving range, and overall efficiency.
- Evaluation at Constant Speeds: The vehicle performance with the proposed control system is evaluated at various constant speeds, analyzing its powertrain efficiency, handling characteristics, and overall responsiveness.
- Evaluation under WLTC Standard Cycle and Real-World Driving Conditions in Tehran Driving Conditions: The performance of the control system is evaluated under the WLTC and real-world driving conditions in the city of Tehran. This comprehensive evaluation assesses the system's ability to optimize energy consumption, driving range, and emissions under diverse driving scenarios.

## EV mode

PHEVs offer a significant advantage of operation in pure electric mode. Before discussing the hybrid modes and the proposed control system, an analysis of the battery pack and vehicle specifications in all-electric mode is essential. The WLTC is used as a benchmark for evaluation. The vehicle speed and SOC versus time during the WLTC cycle are illustrated in Figure 5. It can be observed that 27% of the battery charge is consumed during the WLTC cycle, resulting in a driving range of approximately 23 kilometers. From another perspective, this is equal to an all-electric range of about 84 kilometers, assuming an 80% utilization of the 20-kWh battery capacity.

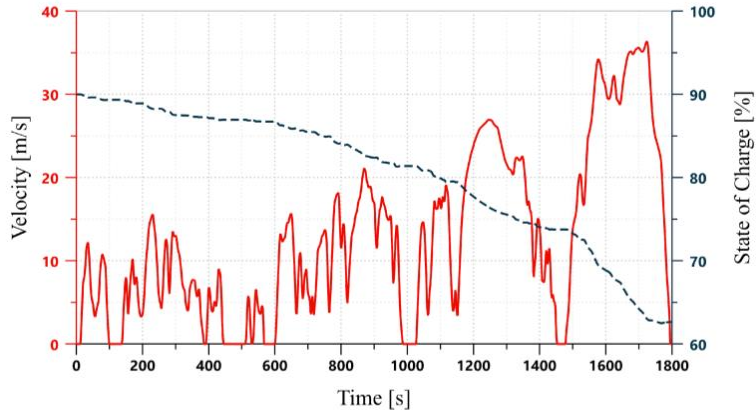

Figure 5: Speed and battery charge level versus time for full-electric mode in the WLTC cycle

## Evaluation of the control system at constant speeds

To gain a deeper understanding of the system's performance, it is crucial to analyze its behavior at constant speeds before examining its performance in driving cycles. In this section, the control system is evaluated for a 600-second simulation at various constant speeds. The changes in battery SOC ($\Delta SOC$), gasoline fuel consumption, and equivalent electric energy consumption at different speeds are shown in Figure 6. The SOC reflects the battery charging or discharging states at different speeds. Positive values indicate battery discharge, while negative values indicate battery charging. For instance, at 140 km/h, the battery discharges by 9.61 %, while at 120 km/h, it charges by -1.98 %. This pattern suggests a key observation: battery discharging is significantly higher at higher speeds, particularly at 140 km/h, while at 120 km/h, the battery is actually charging. This difference could be due to the differential utilization of energy recovery and consumption management systems at different speeds.

Fuel consumption per 100 kilometers varies with speed. Analysis shows that fuel consumption is lower at lower speeds (20 and 40 km/h). At 20 km/h, fuel consumption is 0.744 liters per 100 kilometers, significantly lower than at 140 km/h, where it reaches 5.643 liters per 100 kilometers. This clearly demonstrates that increasing speed leads to higher fuel consumption. Interestingly, fuel consumption at 120 km/h (6.26 liters per 100 kilometers) is higher than at 140 km/h, which could be attributed to battery usage at this speed. This trend is due to the fact that at higher speeds, i.e., more than 80 km/h, the controller prioritizes the use of ICE over battery causing a decrease in battery usage. More specifically, at 120 km/h owing to the optimal engine performance state, ICE works harder and stores the extra energy in the battery for future use. Increasing the speed to 140 km/h, however, results in increased battery usage to maintain the requested speed and lower overall fuel consumption.

The data spanning a 600-second interval from the vehicle stop to reaching a constant speed is shown in Figure 7. For clarity and comparability, CO2 emissions are scaled by $10^6$, while NOx emissions are multiplied by 10. As shown in this figure, the control system prioritizes full electric mode at speeds below 100 km/h, based on predictive battery charge level assessments. Consequently, the ICE remains inactive. As illustrated in Figure 6, this strategy yields impressive fuel efficiency when the battery is charged from an external power grid.

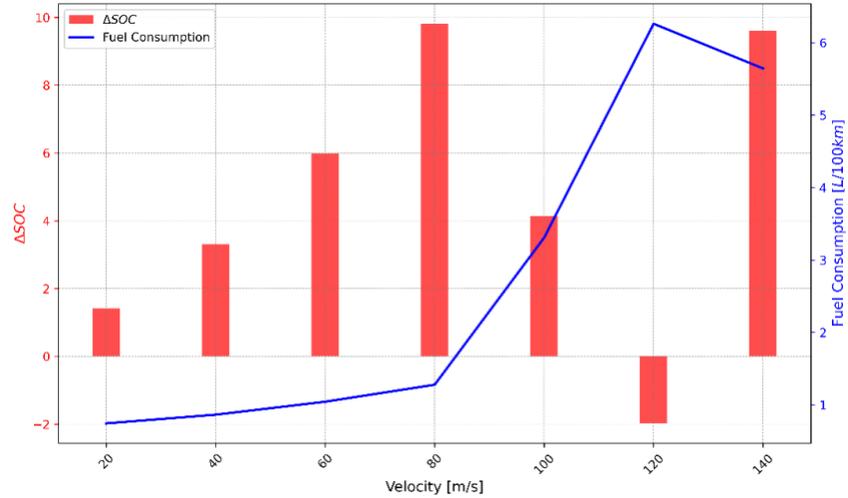

Figure 6: Changes in battery charge level and combined fuel consumption (electric + gasoline) at constant speeds

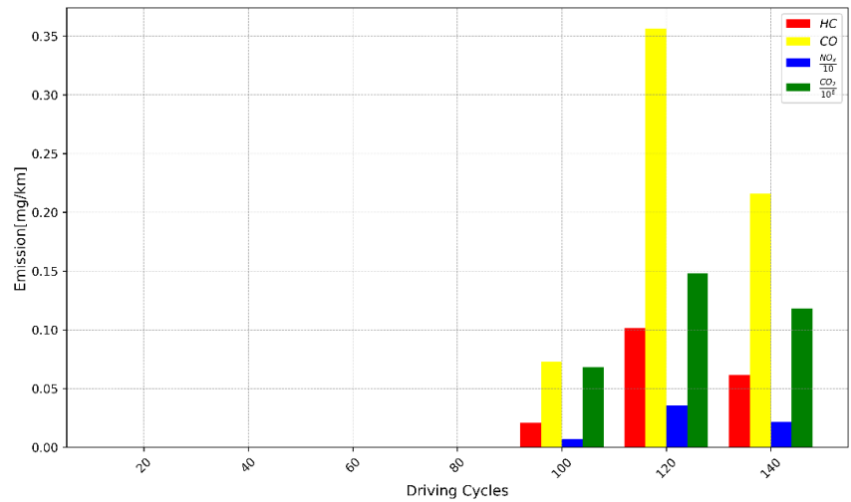

Figure 7: Production of emission at constant speeds for 600 s

However, when speeds exceed 100 km/h, the control system dynamically adjusts its strategy. The SOC is forecasted, and the phase control system ensures a smooth transition from series hybrid to parallel hybrid mode during acceleration, maintaining optimal battery charge levels for future journeys. A comparative analysis of vehicle speeds for 100, 120, and 140 km/h speed reveals significant performance variations. Notably, the 120 km/h speed exhibits superior performance compared to 140 km/h, while 100 km/h lags behind.

Furthermore, higher speeds necessitate hybrid mode engagement, resulting in parameter increments. At 120 km/h, the control system selectively opts for parallel hybrid mode under specific conditions, contributing to a marginally delayed response compared to the series mode at 100 km/h. This precise selection influences the performance points on the fuel consumption map.

### WLTC Driving Cycle Analysis

The assessment of the control system under the WLTC driving cycle, as illustrated in Figure 8 (a), shows its adeptness in managing power allocation between the electric motor and the internal combustion engine (ICE) to provide the required driving power, i.e., $P_d$. Harnessing the insights of the battery forecasting model, which anticipates 30% utilization in all-electric mode, the fuzzy control system adeptly steers towards prioritizing all-electric operation during the initial phases of the cycle. As the vehicle gains momentum and velocity escalates, the control system seamlessly transitions to series hybrid mode, effectively supplementing battery power.

This integrated approach is pivotal in preserving an optimal battery charge level by the end of the journey while concurrently curbing emissions. Delving into the final breakdown of power distribution, the analysis unveils a finely tuned equilibrium, with approximately 51% of the energy sourced from the ICE and the remaining 49% supplied by the battery over the duration of the WLTC driving cycle. This meticulously balanced strategy not only enhances fuel efficiency and diminishes emissions but also safeguards the longevity of the battery pack for sustained performance over time.

The vehicle engine emission profile observed over an 1800-second simulation period for the WLTC driving cycle is shown in Figure 8 (b). Consistently incremental trends in emission levels throughout the simulation, that is attributable to two primary factors, are shown in this figure. Firstly, the depletion of battery charge significantly increases the use of the internal combustion engine (ICE). This augmented reliance on the ICE corresponds with notable escalations in emissions of hydrocarbons (HC), carbon monoxide (CO), and nitrogen oxides (NOx). Particularly striking is the 354% surge in CO2 emissions between the 1000 and 1800-second marks, spotlighting the pronounced impact of highway driving conditions on fuel consumption. Similar upward trajectories are observed for HC (288%), CO (272%), and NOx (390%).

Secondly, the shift to highway driving in the latter phase of the simulation necessitates increased engine power output, leading to amplified fuel consumption and emissions. Conversely, during the initial stages (prior to 1000 seconds), the vehicle predominantly operates in all-electric mode, resulting in minimal emissions. This emission profile reveals the essential influence of driving mode and battery charge level on the pollutant particles. The vehicle driving cycle transition from the low-speed modes to high-speed mode (highway) resulted in the increased reliance on the ICE elevating emissions across all pollutant categories. These findings prove the critical significance of effective battery management strategies and optimized driving behaviors in mitigating real-world emissions.

PHEVs technically require control systems that can adapt and perform efficiently under various driving conditions. A critical aspect of this optimization is understanding the influence of initial battery charge level on vehicle performance and energy consumption. Three initial charge scenarios are considered here, including 90%, 70%, and 50%. The changes in State of Charge (SOC) and equivalent gasoline fuel consumption for these scenarios of the WLTC cycle are compared in Figure 9.

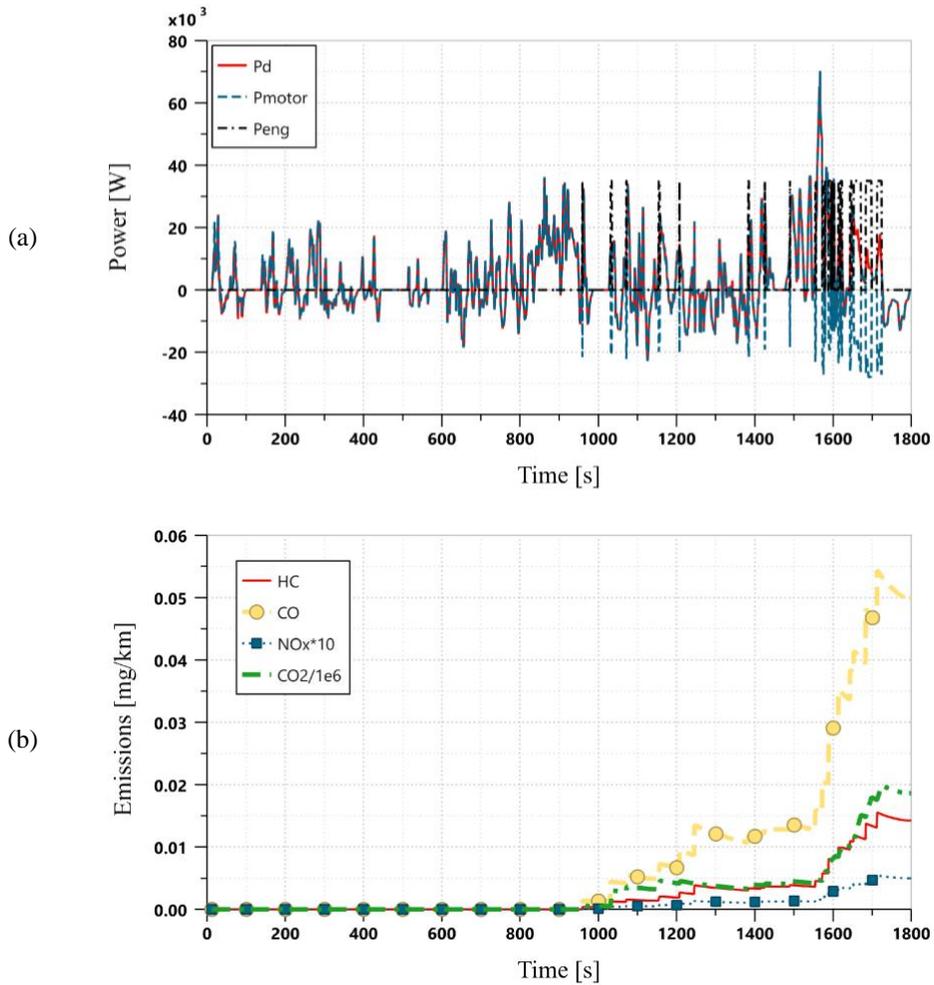

Figure 8: Vehicle performance in the WLTC cycle, (a) Power distribution between the battery and ICE, and (b) The amount of produced emissions

*50% initial charge:* The ICE remains active throughout the driving cycle, with the battery supplementing power during peak demand and charging during lower power phases. The 50% initial charge scenario exhibits the highest fuel consumption, reaching approximately 5 liters per 100 kilometers. Compared to a commercial 2018 Mitsubishi Outlander PHEV in full-electric mode, this represents a 1-liter reduction, highlighting the control system's ability to optimize battery usage. Additionally, 2 liters of engine-generated energy are stored in the battery during the cycle.

*70% initial charge:* The battery prioritizes efficient energy utilization during low-speed driving (up to 1000 seconds), minimizing emissions. Compared to the previous scenario, the 70% scenario follows a similar trend, but the higher initial SOC provides greater flexibility for the control system, leading to a further 1-liter reduction in fuel consumption. Additionally, 0.5 liters equivalent of gasoline are stored in the battery pack during the cycle.

*90% initial charge:* The vehicle operates in all-electric mode until approximately 1000 seconds, maintaining a SOC above 80%. This aligns with the behavior discussed in Figure 8. Subsequently, it transitions to hybrid mode for optimal performance. The lower initial SOC triggers immediate activation of the internal combustion engine (ICE). The ICE operates to ensure the SOC reaches 80% by the end of the cycle.

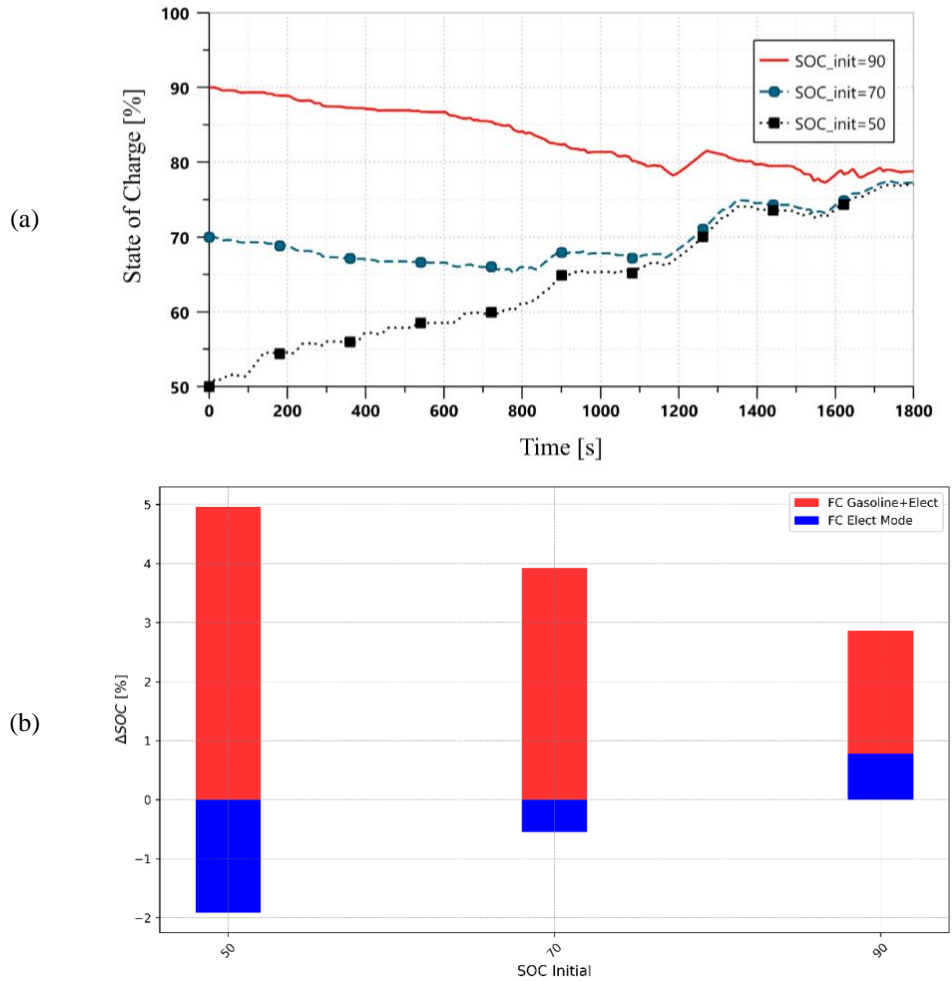

Figure 9: The effect of the initial charging level on (a) SOC, and (b) fuel consumption (L/100km)

### Real-world Driving Cycles Analysis

Driving cycles are standardized test procedures to assess vehicle performance, fuel consumption, and emissions under real recorded conditions. These cycles simulate real-world driving patterns and are crucial for comparing different vehicles' efficiency and environmental impact. Six driving cycles designated, "Tehran 1" through "Tehran 6", are compared in this section to understand their characteristics and suitability for evaluating vehicle performance in various urban and extra-urban scenarios.

As for the driving cycles, different parameters, including total distance, driving time, and average speed are compared as reported in Table 2. Notably, Tehran 5 exhibits the longest total distance (over 128 km) and driving time (nearly 4,500 seconds), representing a prolonged trip with extended periods of vehicle operation. Conversely, Tehran 6 displays the shortest total distance (around 46 kilometers) and driving time (less than 1,500 seconds), potentially reflecting an urban driving scenario with frequent start-stops and lower average speeds (about 34 km/h). Further analysis of stop time and maximum speed reinforces these observations. Tehran 5 has the longest stop-time (almost 1,600 seconds), indicating numerous start-stop situations. On the other hand, Tehran 6 exhibits minimal stop-time (only 38 seconds) and a lower maximum speed (about 80 km/h) compared to other cycles.

The performance of a PHEV control system is directly linked to the driving behavior, which significantly impacts emissions and fuel consumption. Analyzing real-world driving patterns, as depicted in Figure 10, can provide valuable insights into the effectiveness of the intelligent control system. Figure 10 illustrates the relationship between speed variations and battery SOC across the six introduced driving routes, with cycles 1 to 3 on the left and 4 to 6 on the right. A trend where higher average speeds generally correspond to larger SOC changes was observed in these driving cycles. The specific SOC variations for driving cycles Tehran 1 to Tehran 6 are 16.16%, 6.7%, 6.48%, 17.08%, 14.64%, and 6.88%, respectively. This suggests a potential correlation between speed and battery usage.

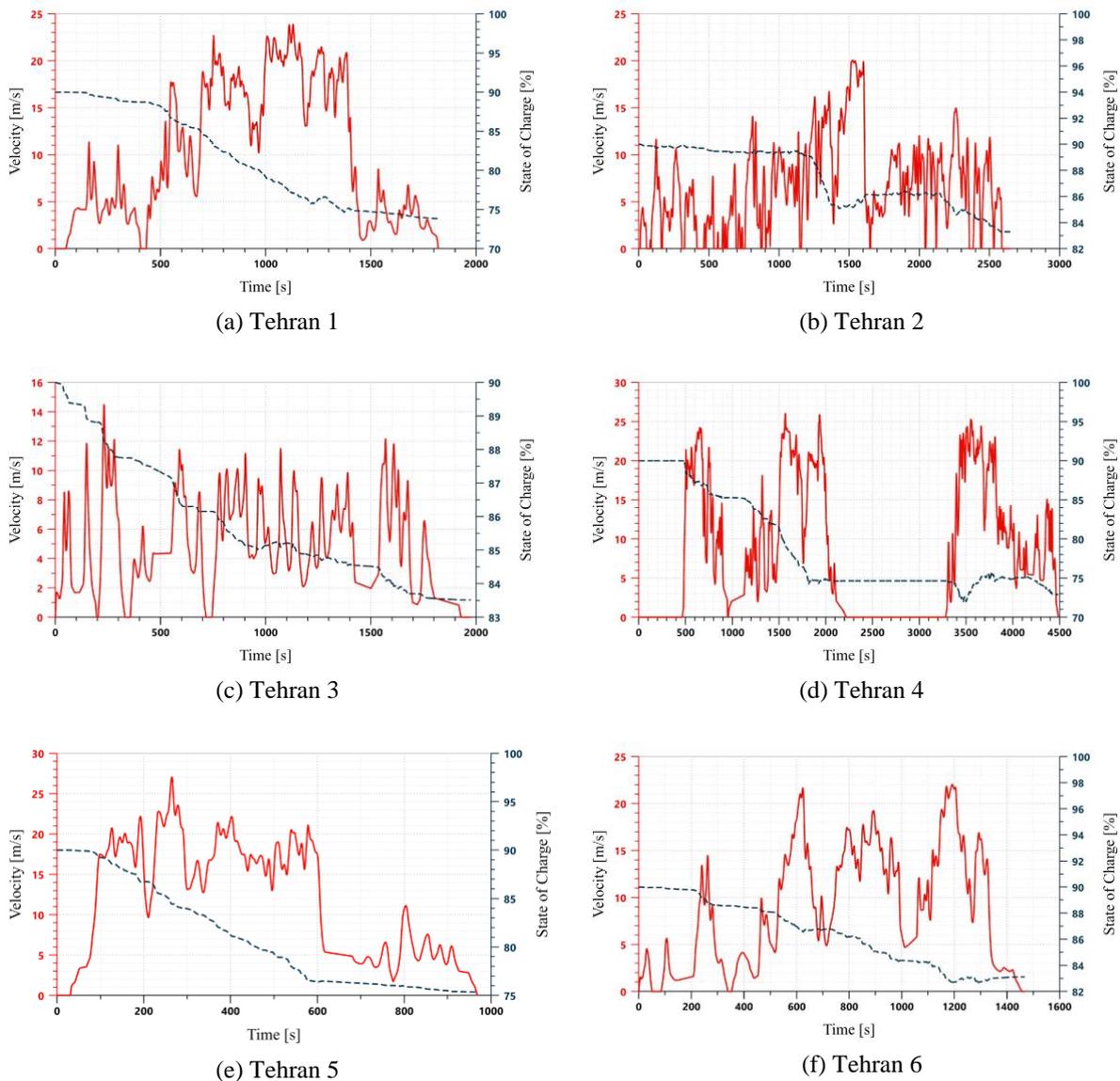

(a) Tehran 1  (b) Tehran 2
(c) Tehran 3  (d) Tehran 4
(e) Tehran 5  (f) Tehran 6

Figure 10: Speed according to time and battery SOC in hybrid mode in real driving cycles from the city of Tehran

Fuel consumption (FC) data for each cycle are shown in Figure 11 (a). In the WLTC standardized cycle, moderate values are observed for both electric FC (0.487) and total FC (1.277). In contrast, other cycles, representing real-world driving scenarios in Tehran, display a broader range of fuel consumption. As such, different fuel consumption rates for both electric and gasoline modes were recorded in these driving cycles. This indicates a balance between electric and gasoline usage, potentially reflecting a mix of urban and extra-urban driving conditions.

Tehran 2 and Tehran 6: These cycles exhibit relatively low FC total values, 0.76 and 0.94, respectively, and lower FC Elect in comparison with other Tehran cycles. This suggests efficient control system operation, potentially due to shorter distances, lower average speeds, and fewer start-stop situations (as observed in the previous analysis, Table 2).

Tehran 1, Tehran 3, Tehran 4, and Tehran 5: These cycles showed higher FC total values (ranging from 2.22 to 3.32) and higher FC Elect compared to Tehran 2 and 6. This indicated a greater reliance on the ICE, likely due to factors like longer distances, potentially higher speeds on extra-urban roads, or more frequent start-stop situations that can limit battery efficiency.

Figure 11 (a) presents the fuel consumption data of a control system that predicts driving behavior based on charge levels. During driving cycles with shorter distances, lower speeds, and fewer stops (Tehran 2 and 6), the system prioritized electric mode, leading to lower gasoline fuel consumption and electricity usage. Conversely, for cycles with longer distances, higher speeds, or more start-stop situations (Tehran 1, 3, 4, and 5), the system operated in ICE mode more frequently, resulting in higher total gasoline fuel consumption and electricity usage. This adaptability demonstrates the control system's effectiveness in optimizing fuel consumption based on real-world driving conditions.

The effects of driving conditions on emission production, considering $CO_2$, HC, CO, and NOx across various driving cycles (WLTC, Tehran 1-6) are shown in Figure 11 (b). The data presented in Figure 11 (b) reveals significant variations in emission levels across the driving cycles. The standardized WLTC cycle exhibited moderate emission values for most pollutants ($CO_2$: 18.64 g/km, HC: 0.014 mg/km, CO: 0.050 mg/km, NOx: 0.0005 mg/km). Conversely, the Tehran cycles, representing real-world driving scenarios in the city of Tehran, showcased a wider range of emissions.

An interesting pattern was achieved in terms of comparing emissions to driving characteristics. Cycles with potentially lower average speeds and fewer stops (e.g., Tehran 2 and 6) exhibited the lowest emission values across all pollutants. Conversely, cycles with potentially higher average speeds or more start-stop situations (e.g., Tehran 1, 4, and 5) showed higher emissions. As for $CO_2$ and NOx, these pollutants are generally produced in higher levels in cycles with potentially higher average speeds or more start-stop situations. This achievement revealed that frequent acceleration and deceleration phases, common in urban driving with start-stops, can contribute to increased $CO_2$ and NOx emissions. As for HC and CO, these pollutants are generally lower in cycles with potentially lower average speeds and fewer stops. This indicated that HC and CO emissions were more prevalent during cold starts or inefficient combustion at higher engine loads, occurring during frequent start-stop conditions or rapid accelerations.

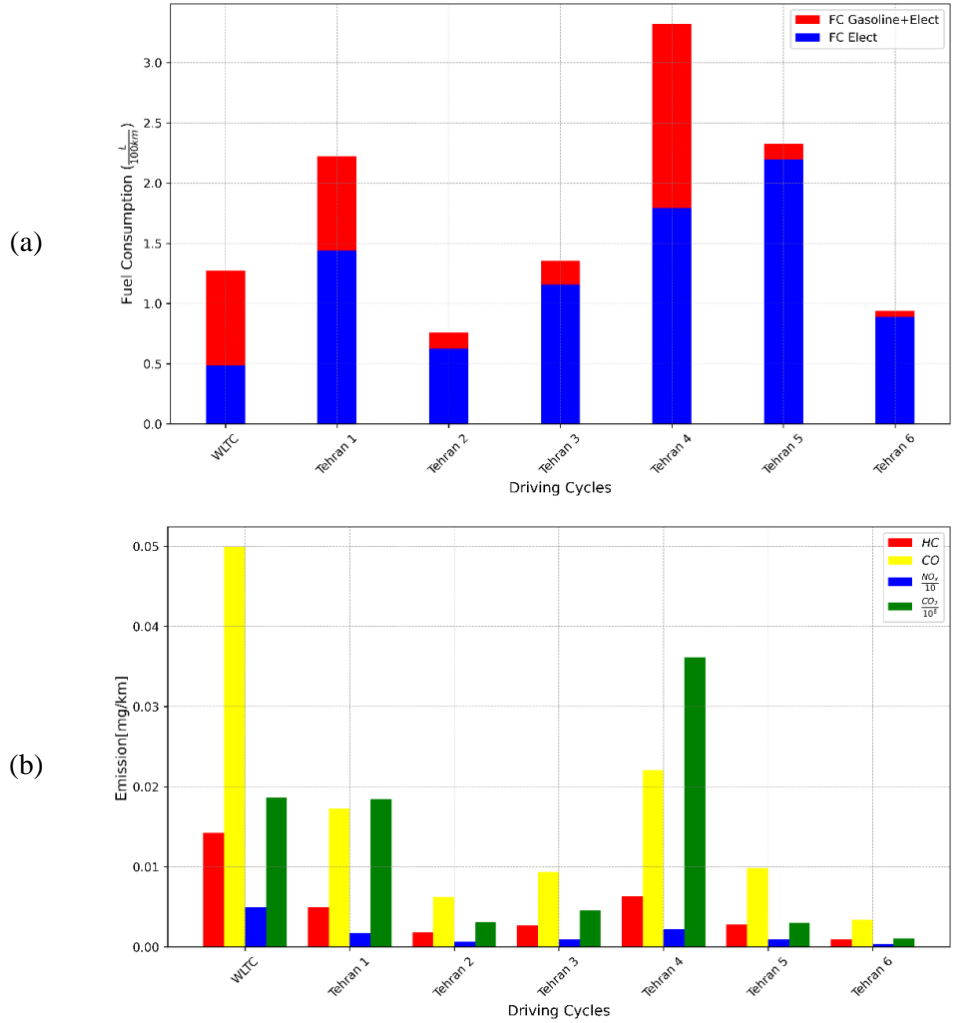

Figure 11: Comparing vehicle performance in different cycles: (a) combined fuel consumption (electricity + gasoline), and (b) amount of emissions

### Performance and state of health

The battery charging and discharging trends significantly impact the battery SOH. Frequent cycling, particularly during driving, can accelerate SOH degradation. Therefore, evaluating the control system performance under different SOH conditions is crucial for optimizing both emissions reduction and battery usage. The battery charge level behavior during the WLTC cycle for two SOH scenarios: 100% and 80%, are shown in Figure 12.

As for the 100% SOH scenario, the control system prioritized electric mode for a longer duration, likely switching to hybrid mode only after a significant drop in battery capacity (10%). This approach minimized reliance on the internal combustion engine, leading to lower emissions. For the 80% SOH scenario, as battery health has deteriorated, the control system adopted a more conservative approach, switching to hybrid mode earlier, potentially after a smaller decrease in battery capacity, i.e., $\Delta SOC < 10\%$. This is also true for any SOH below 100%, with lower SOHs causing an earlier switch to hybrid mode. This indicated that the control system estimated energy consumption based on feedback from previous cycles and adapted the strategy to maintain battery health. While this reduced reliance on EV mode, it was a necessary trade-off to prolong battery life. These findings highlighted the control system's ability to adapt its control strategy based on SOH. Prioritizing EV mode at higher SOH (100%) resulted in lower emissions but accelerated battery degradation. Conversely, a more conservative approach at lower SOH (80%) reduced battery electrical and thermal stress but increased ICE usage and hence associated emissions.

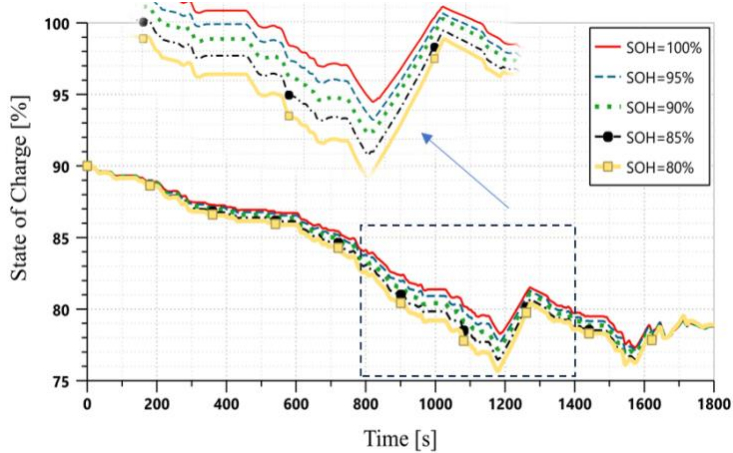

Figure 12: Battery charge level at different battery SOH for the WLTC driving cycle

The fuel consumption values for both electric and hybrid modes ($FC_{elect}$ and $FC_{total}$) across various driving cycles and SOH conditions are shown in Figure 13. For the WLTC cycle, as SOH increased from 80% to 100%, a noticeable increase in $FC_{elect}$ value from 0.6169487 to 0.7852057 L/100km was observed, while $FC_{total}$ decreased from 2.986327 to 2.867019 L/100km. This finding indicated that a healthier battery pack enables more reliance on electric power (full electric mode usage), reducing the total fuel consumption.

In the Tehran 1 cycle, a similar trend was observed where $FC_{elect}$ increased from 1.189768 to 1.438852 L/100km as SOH improved, and $FC_{total}$ decreased from 2.424007 to 2.220519 L/100km, supporting the trend of enhanced electric mode utilization with higher values for SOH. For Tehran 2, $FC_{elect}$ values remain relatively stable across different SOH levels, with a slight increase from 0.6668828 to 0.672369 L/100km. The $FC_{total}$ also showed minor variations, indicating a consistent performance across varying SOH conditions in this specific cycle. In Tehran 3, $FC_{elect}$ varied in the range of 1.109592 L/100km at 80% SOH to 1.159643 L/100km at 100% SOH, with a corresponding decrease of $FC_{total}$ from 1.394845 to 1.353792 L/100km, highlighting improved efficiency with higher values of SOH.

Tehran 4 cycle indicated a more pronounced increase in $FC_{elect}$ from 1.449789 to 1.689414 L/100km, while $FC_{total}$ remained relatively stable with slight fluctuations, showing a complex relationship between electricity and total fuel consumption. For Tehran 5, $FC_{elect}$ values range significantly from 2.018798 to 2.200067 L/100km as SOH improved, with $FC_{total}$ showing a general decreasing trend from 2.350398 to 2.225108 L/100km, emphasizing a higher reliance on electric power with better battery health. Lastly, the Tehran 6 cycle showed stable $FC_{elect}$ values, 0.9197551 L/100km across all SOH levels, with $FC_{total}$ remaining consistent close to 0.9215074 L/100km, indicating the minimal impact of SOH on fuel consumption in this cycle.

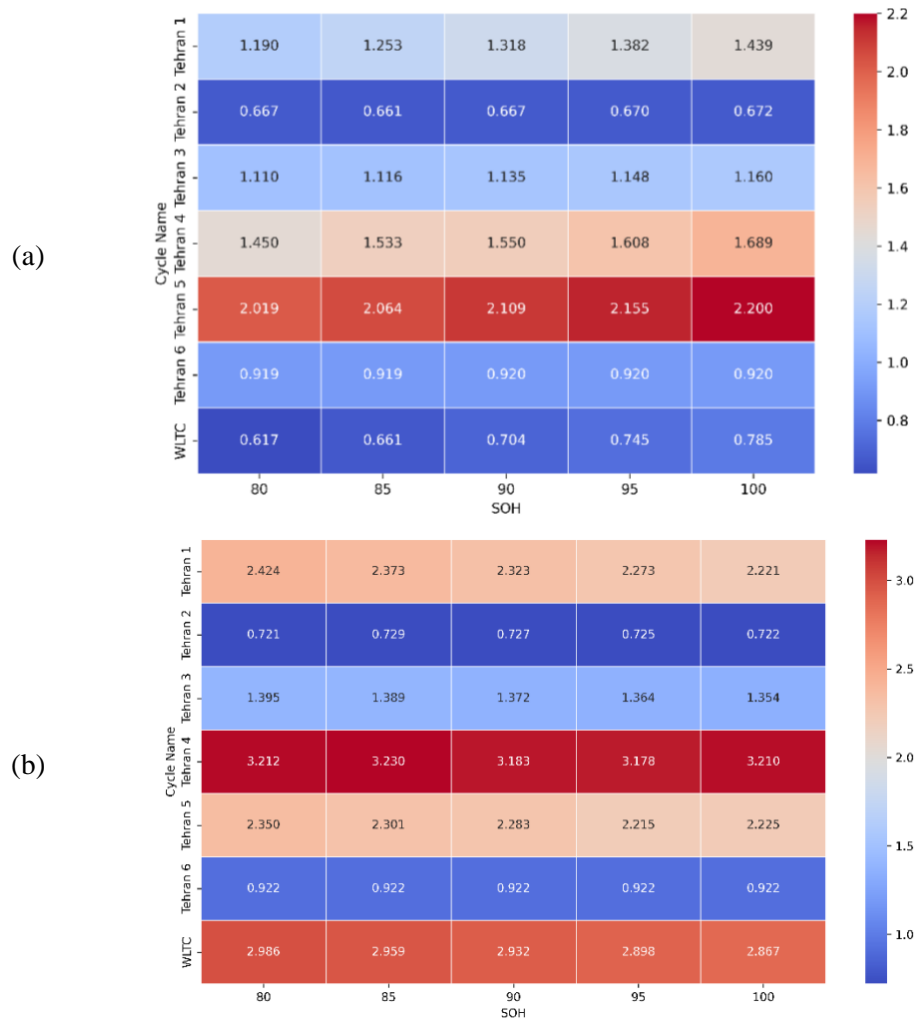

Figure 13: The performance of the control system in different battery states of health for the WLTC cycle and real driving cycles: (a) the amount of electricity consumption equivalent to gasoline in terms of L/100km, and (b) the combined fuel consumption in terms of L/100km

Overall, the data indicated that higher SOH generally leads to increased electricity usage and reduced total fuel consumption, although the extent of this effect varied across different driving cycles. This finding highlighted the importance of maintaining battery health to optimize fuel efficiency and emissions.

## 6. Conclusions

In this study, the performance of a PHEV was improved by developing a smart control system, which was programmed with the help of machine learning tools. Using such a technique, the amount of electricity consumption in the full-electric mode was predicted and the power distribution in different performance modes was investigated and determined. Four operation modes, including full-electric, series hybrid, parallel hybrid, and full-ICE modes were studied. As for the controller, the fuzzy controller system was responsible for deciding the operating mode among these four modes.

To check the performance of the vehicle using this controller, the longitudinal dynamics model of the desired vehicle was developed in AMESime. Utilizing this model, the performance of the vehicle was evaluated in the WLTC cycle and 6 real-world driving cycles obtained in the city of Tehran. The results demonstrated that the all-electric prediction on the target route reduced both the vehicle's variability and energy consumption. Furthermore, vehicle energy consumption is significantly reduced in heavy traffic patterns, as evidenced by the dynamics model of the proposed control system compared to the base controller fuel efficiency. The main results of this study are briefly stated below:

- The proposed control system significantly improved the vehicle energy consumption in all-electric mode, optimizing energy consumption and extending the driving range. During the WLTC cycle, a notable reduction in battery charge usage was observed, translating into an extended all-electric range of approximately 84 kilometers.
- At constant speeds, the control system demonstrated its ability to efficiently manage the powertrain. The vehicle battery discharged at higher speeds and charged at lower speeds, indicating effective utilization of energy recovery and consumption management systems. Fuel consumption analysis revealed lower usage at slower speeds, emphasizing the control system's capability to maintain efficiency across varying conditions.
- The control system adeptly handled the WLTC cycle and real-world driving conditions in Tehran. It dynamically transitioned between electric and hybrid modes, optimizing energy distribution and maintaining optimal battery charge levels. This adaptability ensured a balance between fuel efficiency and reduced emissions, particularly in urban driving scenarios.
- The study highlighted the influence of initial battery charge levels on vehicle performance. Higher initial charge levels allowed for prolonged all-electric operation, reducing fuel consumption and emissions. Conversely, lower initial charges prompted earlier ICE activation to maintain SOC, demonstrating the control system's flexibility in optimizing battery usage based on initial conditions.
- Analysis of real-world driving patterns in Tehran revealed significant variations in fuel consumption and emissions across different driving cycles. The control system prioritized electric propulsion in shorter, slower cycles, leading to lower fuel consumption. In longer, faster cycles, the system utilized the ICE more frequently, highlighting its ability to adapt to diverse driving conditions and optimize performance accordingly.
- The control system performance was evaluated under different battery SOH conditions. Higher SOH facilitated extended use of EV mode, reducing total fuel consumption but potentially accelerating battery degradation. Lower SOH prompted earlier transitions to hybrid mode, balancing battery life and fuel efficiency. This adaptability underscores the importance of maintaining battery health for optimal vehicle performance.

In summary, the proposed control system for PHEVs demonstrated significant improvements in efficiency, adaptability, and overall performance across various driving conditions. By leveraging machine learning tools and a fuzzy decision-making system, the control system effectively optimized power distribution, reduced energy consumption, and minimized emissions. The study's findings emphasize the potential for advanced control systems to enhance the sustainability and efficiency of electric vehicles, contributing to a greener and more efficient transportation future.

# References

[1] Jui JJ, Ahmad MA, Molla MMI, Rashid MIM. Optimal energy management strategies for hybrid electric vehicles: A recent survey of machine learning approaches. J Eng Res 2024. https://doi.org/10.1016/J.JER.2024.01.016.

[2] Zhang F, Hu X, Langari R, Cao D. Energy management strategies of connected HEVs and PHEVs: Recent progress and outlook. Prog Energy Combust Sci 2019;73:235–56. https://doi.org/10.1016/J.PECS.2019.04.002.

[3] Redelbach M, Özdemir ED, Friedrich HE. Optimizing battery sizes of plug-in hybrid and extended range electric vehicles for different user types. Energy Policy 2014;73:158–68. https://doi.org/10.1016/j.enpol.2014.05.052.

[4] Xu L, Ouyang M, Li J, Yang F, Lu L, Hua J. Optimal sizing of plug-in fuel cell electric vehicles using models of vehicle performance and system cost. Appl Energy 2013;103:477–87. https://doi.org/10.1016/j.apenergy.2012.10.010.

[5] Mahmoodi-k M, Montazeri M, Madanipour V. Simultaneous multi-objective optimization of a PHEV power management system and component sizing in real world traffic condition. Energy 2021;233:121111. https://doi.org/10.1016/J.ENERGY.2021.121111.

[6] Song Z, Hofmann H, Li J, Hou J, Zhang X, Ouyang M. The optimization of a hybrid energy storage system at subzero temperatures: Energy management strategy design and battery heating requirement analysis. Appl Energy 2015;159:576–88. https://doi.org/10.1016/j.apenergy.2015.08.120.

[7] Hu Z, Li J, Xu L, Song Z, Fang C, Ouyang M, et al. Multi-objective energy management optimization and parameter sizing for proton exchange membrane hybrid fuel cell vehicles. Energy Convers Manag 2016;129:108–21. https://doi.org/10.1016/j.enconman.2016.09.082.

[8] Qi X, Wu G, Boriboonsomsin K, Barth MJ. Development and Evaluation of an Evolutionary Algorithm-Based OnLine Energy Management System for Plug-In Hybrid Electric Vehicles. IEEE Trans Intell Transp Syst 2017;18:2181–91. https://doi.org/10.1109/TITS.2016.2633542.

[9] Raeesi M, Changizian S, Ahmadi P, Khoshnevisan A. Performance analysis of a degraded PEM fuel cell stack for hydrogen passenger vehicles based on machine learning algorithms in real driving conditions. Energy Convers Manag 2021;248:114793. https://doi.org/10.1016/J.ENCONMAN.2021.114793.

[10] Banvait H, Anwar S, Chen Y. A rule-based energy management strategy for plugin hybrid electric vehicle (PHEV). Proc Am Control Conf 2009:3938–43. https://doi.org/10.1109/ACC.2009.5160242.

[11] Phan D, Bab-Hadiashar A, Fayyazi M, Hoseinnezhad R, Jazar RN, Khayyam H. Interval Type 2 Fuzzy Logic Control for Energy Management of Hybrid Electric Autonomous Vehicles. IEEE Trans Intell Veh 2021;6:210–20. https://doi.org/10.1109/TIV.2020.3011954.

[12] Zhou Q, Zhao D, Shuai B, Li Y, Williams H, Xu H. Knowledge Implementation and Transfer with an Adaptive Learning Network for Real-Time Power Management of the Plug-in Hybrid Vehicle. IEEE Trans Neural Networks Learn Syst 2021;32:5298–308. https://doi.org/10.1109/TNNLS.2021.3093429.

[13] Zhou Q, Li Y, Zhao D, Li J, Williams H, Xu H, et al. Transferable representation modelling for real-time energy management of the plug-in hybrid vehicle based on k-fold fuzzy learning and Gaussian process regression. Appl Energy 2022;305:117853. https://doi.org/10.1016/J.APENERGY.2021.117853.

[14] Song Z, Hofmann H, Li J, Han X, Ouyang M. Optimization for a hybrid energy storage system in electric vehicles using dynamic programing approach. Appl Energy 2015;139:151–62. https://doi.org/10.1016/j.apenergy.2014.11.020.

[15] Xie S, Hu X, Xin Z, Li L. Time-Efficient Stochastic Model Predictive Energy Management for a Plug-In Hybrid Electric Bus with an Adaptive Reference State-of-Charge Advisory. IEEE Trans Veh Technol 2018;67:5671–82. https://doi.org/10.1109/TVT.2018.2798662.

[16] Zhou W, Yang L, Cai Y, Ying T. Dynamic programming for new energy vehicles based on their work modes Part II: Fuel cell electric vehicles. J Power Sources 2018;407:92–104.

https://doi.org/10.1016/j.jpowsour.2018.10.048.

[17] Abd-Elhaleem S, Shoeib W, Sobaih AA. A new power management strategy for plug-in hybrid electric vehicles based on an intelligent controller integrated with CIGPSO algorithm. Energy 2023;265:126153. https://doi.org/10.1016/J.ENERGY.2022.126153.

[18] Sun H, Fu Z, Tao F, Zhu L, Si P. Data-driven reinforcement-learning-based hierarchical energy management strategy for fuel cell/battery/ultracapacitor hybrid electric vehicles. J Power Sources 2020;455. https://doi.org/10.1016/j.jpowsour.2020.227964.

[19] Xie S, Hu X, Qi S, Lang K. An artificial neural network-enhanced energy management strategy for plug-in hybrid electric vehicles. Energy 2018;163:837–48. https://doi.org/10.1016/j.energy.2018.08.139.

[20] Chen Z, Liu Y, Zhang Y, Lei Z, Chen Z, Li G. A neural network-based ECMS for optimized energy management of plug-in hybrid electric vehicles. Energy 2022;243. https://doi.org/10.1016/j.energy.2021.122727.

[21] Chen SY, Hung YH, Wu CH, Huang ST. Optimal energy management of a hybrid electric powertrain system using improved particle swarm optimization. Appl Energy 2015;160:132–45. https://doi.org/10.1016/j.apenergy.2015.09.047.

[22] Chen Z, Xiong R, Cao J. Particle swarm optimization-based optimal power management of plug-in hybrid electric vehicles considering uncertain driving conditions. Energy 2016;96:197–208. https://doi.org/10.1016/j.energy.2015.12.071.

[23] Li G, Zhang J, He H. Battery SOC constraint comparison for predictive energy management of plug-in hybrid electric bus. Appl Energy 2017;194:578–87. https://doi.org/10.1016/j.apenergy.2016.09.071.

[24] Shen P, Zhao Z, Zhan X, Li J. Particle swarm optimization of driving torque demand decision based on fuel economy for plug-in hybrid electric vehicle. Energy 2017;123:89–107. https://doi.org/10.1016/j.energy.2017.01.120.

[25] Wang Z, Jiao X. Optimization of the powertrain and energy management control parameters of a hybrid hydraulic vehicle based on improved multi-objective particle swarm optimization. Eng Optim 2021;53:1835–54. https://doi.org/10.1080/0305215X.2020.1829612.

[26] Lin C-C, Peng H, Grizzle JW. A stochastic control strategy for hybrid electric vehicles, Institute of Electrical and Electronics Engineers (IEEE); 2018, p. 4710–5 vol.5. https://doi.org/10.23919/acc.2004.1384056.

[27] Moura SJ, Fathy HK, Callaway DS, Stein JL. A stochastic optimal control approach for power management in plug-in hybrid electric vehicles. IEEE Trans Control Syst Technol 2011;19:545–55. https://doi.org/10.1109/TCST.2010.2043736.

[28] Li H, Ravey A, N'Diaye A, Djerdir A. Online adaptive equivalent consumption minimization strategy for fuel cell hybrid electric vehicle considering power sources degradation. Energy Convers Manag 2019;192:133–49. https://doi.org/10.1016/J.ENCONMAN.2019.03.090.

[29] Pisu P, Rizzoni G. A comparative study of supervisory control strategies for hybrid electric vehicles. IEEE Trans Control Syst Technol 2007;15:506–18. https://doi.org/10.1109/TCST.2007.894649.

[30] Bashash S, Moura SJ, Forman JC, Fathy HK. Plug-in hybrid electric vehicle charge pattern optimization for energy cost and battery longevity. J Power Sources 2011;196:541–9. https://doi.org/10.1016/J.JPOWSOUR.2010.07.001.

[31] Li P, Jiao X, Li Y. Adaptive real-time energy management control strategy based on fuzzy inference system for plug-in hybrid electric vehicles. Control Eng Pract 2021;107:104703. https://doi.org/10.1016/J.CONENGPRAC.2020.104703.

[32] Akhtar S, Adeel M, Iqbal M, Namoun A, Tufail A, Kim KH. Deep learning methods utilization in electric power systems. Energy Reports 2023;10:2138–51. https://doi.org/10.1016/J.EGYR.2023.09.028.

[33] Zhang Y, Li Q, Wen C, Liu M, Yang X, Xu H, et al. Predictive equivalent consumption minimization strategy based on driving pattern personalized reconstruction. Appl Energy 2024;367:123424. https://doi.org/10.1016/J.APENERGY.2024.123424.

[34] Shojaeefard MH, Mollajafari M, Edalat Pishe N, Mohsen Mousavi S. Plug-in fuel cell vehicle performance and battery sizing optimization based on reduced fuel cell energy consumption and waste heat. Sustain Energy Technol Assessments 2023;56:103099. https://doi.org/10.1016/J.SETA.2023.103099.

[35] Jungst RG, Nagasubramanian G, Case HL, Liaw BY, Urbina A, Paez TL, et al. Accelerated calendar and pulse life analysis of lithium-ion cells. J Power Sources 2003;119–121:870–3. https://doi.org/10.1016/S0378-7753(03)00193-9.

[36] Feng F, Teng S, Liu K, Xie J, Xie Y, Liu B, et al. Co-estimation of lithium-ion battery state of charge and state of temperature based on a hybrid electrochemical-thermal-neural-network model. J Power Sources 2020;455:227935. https://doi.org/10.1016/J.JPOWSOUR.2020.227935.

[37] Mansour S and Raeesi M. Performance assessment of fuel cell and electric vehicles taking into account the fuel cell degradation, battery lifetime, and heating, ventilation, and air conditioning system. Int J Hydrogen Energy 2024;52:834-55. https://doi.org/10.1016/J.IJHYDENE.2023.05.315.

[38] Shojaeefard MH, Raeesi M. Dynamic analysis and performance improvement of a GDI engine and fuel cell under real driving conditions using machine learning technique. Int J Hydrogen Energy 2023. https://doi.org/10.1016/J.IJHYDENE.2023.10.102.

[39] Ahmadi P, Raeesi M, Changizian S, Teimouri A, Khoshnevisan A. Lifecycle assessment of diesel, diesel-electric and hydrogen fuel cell transit buses with fuel cell degradation and battery aging using machine learning techniques. Energy 2022;259:125003. https://doi.org/10.1016/J.ENERGY.2022.125003.

[40] 2021 Outlander PHEV. 2021.

[41] Outlander PHEV Engine - 2.4L 16-Valve Efficiency | Mitsubishi Motors n.d.

[42] Mashadi B, Crolla D. Vehicle powertrain systems. 1st ed. Wiley; 2012.

[43] Ehsani, M., Gao, Y., Longo, S., & Ebrahimi K. Modern Electric, Hybrid Electric, and Fuel Cell Vehicles, Third Edition. 3rd ed. CRC Press; 2018. https://doi.org/10.1201/9780429504884.

[44] Hu X, Zheng Y, Howey DA, Perez H, Foley A, Pecht M. Battery warm-up methodologies at subzero temperatures for automotive applications: Recent advances and perspectives. Prog Energy Combust Sci 2020;77:100806. https://doi.org/10.1016/j.pecs.2019.100806.

[45] Di Domenico D, Prada E, Creff Y. An Adaptive Strategy for Li-ion Battery SOC Estimation. IFAC Proc Vol 2011;44:9721–6. https://doi.org/10.3182/20110828-6-IT-1002.01119.

[46] Mansour S, Jalali A, Ashjaee M, and Houshfar E. Multi-objective optimization of a sandwich rectangular-channel liquid cooling plate battery thermal management system: A deep-learning approach. Energy Convers Manag 2023;290:117200. https://doi.org/10.1016/J.ENCONMAN.2023.117200.

[47] Timilsina L, Badr PR, Hoang PH, Ozkan G, Papari B, and Edrington CS. Battery Degradation in Electric and Hybrid Electric Vehicles: A Survey Study. IEEE Access 2023. http://doi.org/10.1109/ACCESS.2023.3271287.

[48] Ge MF, Liu Y, Jiang X, Liu J. A review on state of health estimations and remaining useful life prognostics of lithium-ion batteries. Meas J Int Meas Confed 2021;174:109057. https://doi.org/10.1016/j.measurement.2021.109057.

[49] Li Y, Stroe DI, Cheng Y, Sheng H, Sui X, Teodorescu R. On the feature selection for battery state of health estimation based on charging–discharging profiles. J Energy Storage 2021;33:102122. https://doi.org/10.1016/j.est.2020.102122.